\documentclass[12pt]{article}
\usepackage{amsmath,amsfonts,amsthm}
\usepackage{mathtools}
\usepackage{geometry}

\newtheorem{thm}{Theorem}[section]
\newtheorem{lemma}[thm]{Lemma}
\newtheorem{prop}[thm]{Proposition}

\theoremstyle{definition}
\newtheorem{defi}{Definition}

\theoremstyle{remark}
\newtheorem{rem}[thm]{Remark}

\newcommand{\vertiii}[1]{{\left\vert\kern-0.25ex\left\vert\kern-0.25ex\left\vert #1 
    \right\vert\kern-0.25ex\right\vert\kern-0.25ex\right\vert}}
\def\Me{\mathcal M}

\def\Ne{\mathcal N}
\def\Ha{\mathcal H}
\def\Ka{\mathcal K}
\def\Fe{\mathcal F}

\def\Tr{\mathrm{Tr}\,}
\def\states{\mathfrak S}
\def\<{\langle}
\def\>{\rangle}

\begin{document}
\title{R\'enyi relative entropies and noncommutative $L_p$-spaces II}
\author{Anna Jen\v cov\'a \thanks{jenca@mat.savba.sk}\\ \small \emph{Mathematical Institute, Slovak Academy of Sciences}\\
\small \emph{ \v Stef\'anikova 49, 814 73 Bratislava, Slovakia}}
\date{}

\maketitle
\abstract{ We study an
extension of the sandwiched R\'enyi relative entropies for normal positive functionals on a von Neumann algebra, for
parameter values $\alpha\in [1/2,1)$. This work is intended as a continuation of (A. Jen\v cov\'a, Ann. Henri Poincar\'e 19,
2513-2542, 2018), where the values $\alpha>1$ were studied. We use the Araki-Masuda divergences of [Berta et al.,  Ann.
Henri Poincar\'e 9, 1843-1867, 2018] and  treat them in the framework of Kosaki's noncommutative $L_p$-spaces. 
Using the variational formula, recently obtained by F. Hiai, for $\alpha\in [1/2,1)$, we prove the data processing inequality with respect to positive trace preserving maps and show that for $\alpha\in (1/2,1)$, equality characterizes  sufficiency (reversibility) for any 2-positive trace preserving map.}

\section{Introduction}

In \cite{jencova2018renyi}, we introduced and studied an extension $\tilde D_\alpha $ of the sandwiched R\'enyi relative 
entropy  for $\alpha>1$, from
density matrices to positive normal functionals on a von Neumann algebra $\Me$. These quantities were defined using the
noncommutative $L_p$-spaces due to Kosaki \cite{kosaki1984applications}. We proved a number of properties 
 of $\tilde D_\alpha$, in particular the data processing inequality (DPI) with respect to any positive trace preserving
map $\Phi: L_1(\Me)\to L_1(\Ne)$. Moreover, we proved that if $\Phi$ is 2-positive then equality in DPI is equivalent to
the fact that $\Phi$ is sufficient with respect to $\{\psi,\varphi\}$. 

A similar extension was obtained
in \cite{berta2018renyi} for the larger interval of parameters $\alpha\in [1/2,1)\cup (1,\infty]$. 
These quantities were called the Araki-Masuda divergences because the definition is based on the noncommutative
$L_p$-spaces due to Araki and Masuda \cite{araki1982positive}. 
We showed that for $\alpha>1$, these quantities coincide with $\tilde D_\alpha$. 

More recently, an extension to all $\alpha\in (0,1)\cup (1,\infty)$ was proposed in \cite{gu2019interpolation}, using
interpolation of quasi Banach spaces. Recall  that as shown in \cite{mullerlennert2013onquantum}, DPI cannot hold for $\alpha<1/2$.

The aim of the present work is to continue  \cite{jencova2018renyi} by the study of $\tilde D_\alpha$ 
also for  the parameter values  $\alpha\in [1/2,1)$. We use the Araki-Masuda divergences of
\cite{berta2018renyi}, but we  show that these can be obtained using Kosaki's right $L_p$-spaces, for all $\alpha\in
[1/2,\infty]\setminus \{1\}$. Most of the properties denoted by (a)-(h) in \cite{jencova2018renyi} for all these values,
 or some weaker  versions for  $\alpha\in [1/2,1)$, were proved in \cite{berta2018renyi} and
\cite{hiai2020quantum} (see \cite[Theorem 3.15]{hiai2020quantum}). In particular, the DPI with respect to quantum
channels was proved for $\alpha\in [1/2,1)$. Also, a variational expression was obtained for 
$\alpha\in (0,1)$ in \cite[Lemma 3.18]{hiai2020quantum}, generalizing  the expression obtained in \cite{frank2013monotonicity}  to the setting of von Neumann algebras. 

We complete these results as follows. We add the lower bound in property (d) (describing the relation to the standard
R\'enyi relative entropy). We also complete the variational formula of \cite[Lemma
3.18]{hiai2020quantum} to $\alpha>1$. Moreover, using the variational formula, we prove that also for $\alpha\in [1/2,1)$,
DPI holds for all \emph{positive} trace preserving maps, which seems new even in the finite dimensional case.

We also study  equality in DPI. We prove  that as for $\alpha>1$ (\cite[Theorem 4.6]{jencova2018renyi}), equality in DPI for $\tilde D_\alpha$, $\alpha\in (1/2,1)$ implies sufficiency  for any 2-positive trace preserving map.

This paper is intended as a continuation of \cite{jencova2018renyi}, so all notations, definitions  and results will be 
used without separate introduction. We begin in Section \ref{sec:lp} by showing how the weighted $L_p$-norms of \cite{berta2018renyi} can be  introduced in the
standard form $(\lambda,L_2(\Me),J, L_2(\Me)^+)$
 by using Kosaki's right $L_p$-spaces with respect to a faithful positive normal functional. In Section
\ref{sec:renyi} we  
 introduce $\tilde D_\alpha$ for $\alpha\in [1/2,1)$ and extend the variational formula to $\alpha>1$. 
 In section \ref{sec:dpi}, we prove the DPI with respect to 
 positive trace preserving maps. Section \ref{sec:suffic} is devoted to DPI equality conditions.

\section{Interpolation $L_p$-spaces for von a Neumann algebra} \label{sec:lp}

Let $\varphi_0\in \Me_*^+$ be faithful. We begin with the definition of  the right $L_p$-space with respect to
$\varphi_0$ as introduced  by Kosaki \cite{kosaki1984applications}. Let us consider the continuous embedding $\Me\to L_1(\Me)$,
defined by 
\[
x\mapsto h_{\varphi_0}x, \qquad x\in \Me.
\]
The range $h_{\varphi_0}\Me\subset L_1(\Me)$ of this embedding, endowed with the norm $\|h_{\varphi_0}x\|_{\infty,\varphi_0}^R:=\|x\|$, 
will be denoted by $\Me^R$. For $1\le p\le \infty$, 
$L_p^R(\Me,\varphi_0)$ is defined as  the interpolation space
\[
L_p^R(\Me,\varphi):=C_p(\Me^R,L_1(\Me)).
\]  
We denote the norm in $L_p^R(\Me,\varphi_0)$ by $\|\cdot\|_{p,\varphi_0}^R$. 
According to \cite{kosaki1984applications}, $L_p^R(\Me,\varphi_0)\subseteq L_1(\Me)$ is the (dense) subspace of
operators of the form
\[
h=h^{1/q}_{\varphi_0}k,\ k\in L_p(\Me),\qquad \|h\|_{p,\varphi_0}^R=\|k\|_p.
\]
In particular, $L_\infty^R(\Me,\varphi_0)\simeq \Me^R\simeq \Me$ and $L_1^R(\Me,\varphi_0)= L_1(\Me)$.
Moreover, if $1<p<\infty$, $h= h^{1/q}_{\varphi_0}k$ and  $k=h_\mu^{1/p}u$ is the polar decomposition of $k$ in
$L_p(\Me)$, then  
\begin{equation}\label{eq:fR}
f^R_{p,h,\varphi_0}(z)=\mu(1)^{1/p-z} h_{\varphi_0}^{1-z}h_\mu^zu,\qquad z\in S,
\end{equation}
is a function in  $\mathcal F(\Me^R,L_1(\Me))$ such that
\[
f^R_{p,h,\varphi_0}(1/p)=h,\qquad \|h\|_{p,\varphi_0}^R=\vertiii{f}_{\Fe},
\]
see \cite[Appendix B]{jencova2018renyi}.

For $1< p\le \infty$ and $1/p+1/q=1$, we have for the Banach space dual $L_q(\Me,\varphi_0)\simeq L_p(\Me,\varphi_0)^*$,
 with duality given by
\[
\<h^{1/q}_{\varphi_0}k,h^{1/p}_{\varphi_0}l\>^R=\Tr kl,\qquad k\in L_p(\Me), \ l\in L_q(\Me).
\]

\subsection{Interpolation norms in $L_2(\Me)$}

We next introduce a family of interpolation (semi)norms  in $L_2(\Me)$. 
Let $\varphi\in \Me_*^+$  with $e:=s(\varphi)$ be arbitrary but fixed throughout. Let us also fix some
$\sigma\in \Me_*^+$ such that $s(\sigma)=1-e$, so that $\varphi_0:=\varphi+\sigma$ is faithful. For $\xi\in L_2(\Me)$
and  $1\le p\le \infty$, we define:
\[
\|\xi\|_{p,\varphi}^{(2)}= \begin{dcases} \|h_\varphi^{1/2}\xi\|_{p,\varphi_0}^R & \text{if }  1\le p< 2 \text{ or } \xi=e\xi\\
\infty & \text{otherwise}
\end{dcases}
\]
(with the understanding that $\|h_\varphi^{1/2}\xi\|_{p,\varphi_0}^R=\infty$ if 
$h_\varphi^{1/2}\xi\notin L^R_p(\Me,\varphi_0)$). 

It is immediate from this definition that 
if  $\xi,\eta\in L_2(\Me)$ are such that $\omega_\eta=\omega_\xi$, then we have
 $\|\xi\|_{p,\varphi}^{(2)}=\|\eta\|_{p,\varphi}^{(2)}$, for any $1\le p\le \infty$. Indeed, the condition means that there is a partial isometry
$u\in \Me$ such that $\eta=\xi u$ and $\xi=\eta u^*$. If $h_\varphi^{1/2}\xi=h_{\varphi_0}^{1/q}k$ with $k\in L_p(\Me)$,
then $h_\varphi^{1/2}\eta=h_{\varphi_0}^{1/q}ku$ with $ku\in L_p(\Me)$, so that 
\[
\|\eta\|_{p,\varphi}^{(2)}=\|ku\|_p\le \|k\|_p\|u\|\le \|k\|_p=\|\xi\|_{p,\varphi}^{(2)}.
\]
Similarly, we obtain $\|\xi\|_{p,\varphi}^{(2)}\le \|\eta\|_{p,\varphi}^{(2)}$.

Notice that for $p\ge 2$, $\|\cdot\|_{p,\varphi}^{(2)}$ is a
norm in the subspace  where it is finite and the proposition below shows that this subspace is dense in  $eL_2(\Me)$ and 
closed  in $\|\cdot\|_{p,\varphi}$. For $1\le p<2$, we will see that $\|\cdot\|_{p,\varphi}^{(2)}$ is always finite and
defines a seminorm in $L_2(\Me)$ which is a norm if and only if $\varphi$ is faithful. 

\begin{prop}\label{prop:polar2} 
\begin{enumerate}
\item[(i)]   Let $2\le p\le \infty$. Then 
$\|\xi\|_{p,\varphi}^{(2)}<\infty$ if and only if 
$\xi=h_\varphi^{1/2-1/p}k$
for some $k\in L_p(\Me)$ with $k=ek$. Moreover, such $k$ is unique and we have  $\|\xi\|_{p,\varphi}^{(2)}=\|k\|_p$.

\item[(ii)] Let $1\le p< 2$ and let $k=h_\varphi^{1/p-1/2}\xi$. Then $k\in L_p(\Me)$ and 
$\|\xi\|^{(2)}_{p,\varphi}=  \|k\|_p<\infty$.

\end{enumerate}

\end{prop}

\begin{proof} 
 (i) Assume $\|\xi\|_{p,\varphi}^{(2)}<\infty$, then we must have $e\xi=\xi$ and  
$h_{\varphi}^{1/2}\xi\in L_p^R(\Me,\varphi_0)$. By the polar decomposition in $L_p^R(\Me,\varphi_0)$, there is a
unique element $k\in L_p(\Me)$ such that $h_{\varphi}^{1/2}\xi=h_{\varphi_0}^{1/q}k$, and we have
$\|h_{\varphi}^{1/2}\xi\|^R_{p,\varphi_0}=\|k\|_p$. Since $h_{\varphi_0}^z=h_{\varphi}^z+h_\sigma^z$ for all $z\in
\mathbb C$, we have $eh_{\varphi_0}^z=h_{\varphi_0}^ze=h_{\varphi}^z$, so that
$h_{\varphi}^{1/2}\xi=h_{\varphi_0}^{1/2}\xi$.
Since $\varphi_0$ is faithful, it follows that
\[
\xi=h_{\varphi_0}^{1/2-1/p}k.
\]
Similarly,  we obtain
\[
 h_{\varphi_0}^{1/2-1/p}k=\xi=e\xi= h_{\varphi}^{1/2-1/p}k= h_{\varphi_0}^{1/2-1/p}ek
\]
and  this implies $k=ek$. Conversely, assume that $\xi$ has a decomposition as required,
then clearly $e\xi=\xi$ and 
\[
h_{\varphi_0}^{1/2}\xi=  h_\varphi^{1/q}k=  h_{\varphi_0}^{1/q}k.
\]
The statement now follows by the polar decomposition in $L^R(\Me,\varphi_0)$ and its uniqueness.

(ii)  Assume $1\le p<2$, then $1/p-1/2=1/2-1/q>0$ and  since  $\xi\in L_2(\Me)$, we have $k=h_\varphi^{1/p-1/2}\xi\in L_p(\Me)$. Further,
\[
h_\varphi^{1/2}\xi=h_{\varphi_0}^{1/2}e\xi=h_{\varphi_0}^{1/q}h_{\varphi_0}^{1/2-1/q}e\xi=h_{\varphi_0}^{1/q}k
\]
so that $h_\varphi^{1/2}\xi\in L_p^R(\Me,\varphi_0)$ and $\|h_\varphi^{1/2}\xi\|_{p,\varphi_0}^R=\|k\|_p$.

\end{proof}

From the (right) polar decomposition in $L_p(\Me)$, we obtain:

\begin{prop}[Polar decomposition] Let $\xi\in L_2(\Me)$ and $1<p<\infty$. Then $\|\xi\|_{p,\varphi}^{(2)}<\infty$
 if and only if 
there is some $\mu\in \Me_*^+$ and a partial isometry $u\in \Me$, where  $uu^*=s(\mu)\le e$ and $u^*u=\mathrm{supp}(e\xi)$, such that
\begin{enumerate}
\item[(i)] If $2\le p<\infty$, $\xi= h_\varphi^{1/2-1/p}h_\mu^{1/p}u$;
\item[(ii)] If $1\le p<2$, $h_\varphi^{1/p-1/2}\xi=h_\mu^{1/p}u$.
\end{enumerate}
Moreover, such $\mu$ and $u$ are unique and we have $\|\xi\|_{p,\varphi}^{(2)}=\mu(1)^{1/p}$.

\end{prop}

In the situation of the above proposition,  we will say that $\xi$ has the $p$-polar decomposition $\xi=\mu^{1/p}u$
(with respect to $\varphi$).

\begin{prop}[Duality]\label{prop:duality} Let $1\le p\le \infty$, $1/p+1/q=1$, $\xi,\eta\in L_2(\Me)$. Then
\begin{enumerate} 
\item[(i)] $|(\xi,\eta)|\le \|\xi\|_{p,\varphi}^{(2)}\|\eta\|_{q,\varphi}^{(2)}$;
\item[(ii)] if $\xi=e\xi$ or $1\le p<2$:
\[
\|\xi\|_{p,\varphi}^{(2)}=\sup_{\eta\in L_2(\Me), \|\eta\|_{q,\varphi}^{(2)}\le 1} |(\xi,\eta)|.
\]
\item[(iii)] if $1\le p<2$ and $\xi=\mu^{1/p}u$ is the $p$-polar decomposition of $\xi$, then there is a unique 
element $\tilde \xi\in L_2(\Me)$ with $\|\tilde \xi\|_{q,\varphi}^{(2)}=1$ such that the supremum in (ii) is attained.
This element has the $q$-polar decomposition $\tilde\xi=(\mu(1)^{-1}\mu)^{1/q}u$.

\end{enumerate}

\end{prop}

\begin{proof}  We clearly may suppose that both norms on the right hand side of (i) are finite. Assume that, say $p\ge
2$ and let $k\in L_p(\Me)$ be such that $\xi=h_\varphi^{1/2-1/p}k$. Then 
\[
|(\xi,\eta)|=|\Tr \eta^*h_\varphi^{1/2-1/p}k|\le
\|k\|_p\|h_\varphi^{1/2-1/p}\eta\|_q=\|\xi\|_{p,\varphi}^{(2)}\|\eta\|_{q,\varphi}^{(2)}.
\]
From $k=ek$ we also have
\[
|(\xi,\eta)|=|\Tr \eta^*h_\varphi^{1/2-1/p}k|=|\Tr \eta^*h_{\varphi_0}^{1/2-1/p}ek|.
\]
Since the elements $\zeta h_{\varphi_0}^{1/2-1/p}$ with $\zeta\in L_2(\Me)$ are dense in $L_q(\Me)$ and
\[
\|k\|_p=\sup_{l=le\in L_q(\Me), \|l\|_q\le 1}|\Tr lk|,
\] we obtain (ii) in the case that $p\ge 2$ and 
$\|\xi\|_{p,\varphi}^{(2)}<\infty$. Similarly, one can see that if $\xi=e\xi$ and the supremum in (ii) is finite,
 then the map $h_{\varphi_0}^{1/2}\eta\mapsto (\xi,\eta)$ extends to a bounded linear functional on
$L^R_q(\Me,\varphi_0)$, hence there is some element $h=h_{\varphi_0}^{1/q}k\in L^R_p(\Me,\varphi_0)$ such that
for $\eta\in L_2(\Me)$,
\[
(\xi,\eta)=(\xi,e\eta)=\<h_{\varphi_0}^{1/q}k,h_{\varphi_0}^{1/2}e\eta\>^R_{\varphi_0} =\Tr k
h_\varphi^{1/2-1/p}\eta=(h_\varphi^{1/2-1/p}k^*,\eta)
\]
so that $\xi=h_\varphi^{1/2-1/p}k^*$ and $\|\xi\|_{p,\varphi}^{(2)}<\infty$. This finishes the proof of (ii) in the case
 $p\ge 2$ and $\xi=e\xi$.

Assume next that $p<2$ and let $k=h_\varphi^{1/p-1/2}\xi$. Then since $k=ek$,
\begin{align*}
\|\xi\|_{p,\varphi}^{(2)}&=\|k\|_p=\sup_{l\in L_q(\Me), \|l\|_q\le 1}|\Tr lk|= \sup_{l=le\in L_q(\Me),
\|l\|_q\le 1}|\Tr lk|\\
 &=  \sup_{l=le\in L_q(\Me), \|l\|_q\le 1} |(\xi, h_\varphi^{1/p-1/2}l^*)|=\sup_{\eta\in L_2(\Me),
\|\eta\|_{q,\varphi}^{(2)}\le 1}|(\xi,\eta)|. 
\end{align*}
The statement (iii) follows by the duality of the $L_p(\Me)$ spaces.

\end{proof}

\subsection{Relation to the weighted $L_p$-norms of Berta-Scholz-Tomamichel}

We next recall the definition of the weighted $L_p$-norms in \cite{berta2018renyi} and show that these are equal to the 
$\|\cdot\|^{(2)}_{p,\varphi}$ for the standard representation on $L_2(\Me)$.

Let $\varphi\in \Me_*^+$ and let  $\pi:\Me\to B(\Ha)$ be any $*$-representation on a complex Hilbert space $\Ha$ with
inner product $\<\cdot,\cdot\>_\Ha$. For $\xi\in \mathcal H$, let $\omega_\xi$ be the functional  given by $\xi$, that
is $\omega_\xi(a)= \<\pi(a)\xi,\xi \>_\Ha$. We also denote by $\omega'_\xi$ the corresponding functional on the
commutant: $\omega'_\xi(a')=\<a'\xi, \xi\>_\Ha$, $a'\in \pi(\Me)'$. Let $\Delta(\xi/\varphi)$ denote the spatial derivative as defined in \cite[Section 2.2]{berta2018renyi} (see also \cite[Appendix A.2]{jencova2018renyi}).
The $\varphi$-weighted $p$-norm of $\xi\in \mathcal H$ is defined as:
\begin{enumerate} 
\item for $2\le p\le  \infty$, we define
\[
\|\xi\|_{p,\varphi}^{BST}:= \sup_{\zeta\in \mathcal H, \|\zeta\|=1}\|\Delta(\zeta/\varphi)^{1/2-1/p}\xi\|
\]
if $s(\omega_\xi)\le s(\varphi)$ and $+\infty$ otherwise. Note that the supremum can be infinite also if the condition on the supports holds.
\item for $1\le p<2$, we define
\[
\|\xi\|_{p,\varphi}^{BST}:=\inf_{\zeta\in \mathcal H, \|\zeta\|=1, s(\omega'_\zeta)\ge s(\omega'_\xi)} \|\Delta(\zeta/\varphi)^{1/2-1/p}\xi\|.
\]

\end{enumerate}

According to \cite{berta2018renyi}, this quantity  depends only on the functionals $\varphi$ and $\omega_\xi$ and not on
the representation $\pi$ or the representing vector $\xi$. Moreover, similar duality relation holds as those in Proposition \ref{prop:duality}, in particular for any representation 
$\pi:\Me\to B(\Ha)$ and any $\xi,\eta\in \Ha$,
\begin{equation}\label{eq:duality_BST}
 |\<\xi,\eta\>_\Ha|\le \|\xi\|_{p,\varphi}^{BST}\|\eta\|_{q,\varphi}^{BST}.
\end{equation}

Let us now  assume that $\Ha=L_2(\Me)$ and 
 $\pi=\lambda: \Me\to B(L_2(\Me))$ is the representation by left multiplication.
By \cite[Appendix A.2]{jencova2018renyi} (notice a small mistake there), we have for $\eta\in L_2(\Me)$
\[
\Delta(\eta/\varphi)=F^*_{\eta,h_\varphi^{1/2}}\bar F_{\eta,h_\varphi^{1/2}}=J\Delta_{\omega,\varphi}J,
\]
where  $\omega=\omega_{\eta^*}$.
 It follows that for all   $\xi\in L_2(\Me)$, we have 
\begin{align*}
\|\xi\|_{p,\varphi}^{BST}&= \begin{dcases} \sup_{\omega\in \states_*(\Me)}\|\Delta_{\omega,\varphi}^{1/2-1/p}\xi^*\|_2 &
\text{if } s(\omega_\xi)\le s(\varphi)\\
                 +\infty & \text{otherwise} \end{dcases},&  2&\le p\le\infty\\
\|\xi\|_{p,\varphi}^{BST}&=\inf_{\omega\in \states_*(\Me), s(\omega)\ge s(\omega_{\xi^*})}
\|\Delta_{\omega,\varphi}^{1/2-1/p}\xi^*\|_2,&  1&\le p< 2,
\end{align*}
(recall that $\states_*(\Me)$ denotes the set of states of $\Me$).

\begin{prop}\label{prop:bst}  Let $\xi\in L_2(\Me)$, $\omega=\omega_\xi$. We have 
\[
\|\xi\|_{p,\varphi}^{BST}=\|\xi\|_{p,\varphi}^{(2)}=\|h_\omega^{1/2}\|_{p,\varphi}^{(2)}.
\]

\end{prop}

\begin{proof} Since both norms depend only on $\omega$, we may suppose that $\xi=h_\omega^{1/2}$. 
 Assume first that $1\le p<2$. 
Using the properties of the relative modular operator (\cite[Appendix A.1]{jencova2018renyi}), we see that 
\begin{align*}
\|h_\omega^{1/2}\|_{p,\varphi}^{BST}&=\inf_{\psi\in \states_*(\Me), s(\psi)\ge s(\omega)}
\|\Delta_{\psi,\varphi}^{1/2-1/p}h_\omega^{1/2}\|_2\\
&=\inf_{\psi\in \states_*(\Me), s(\psi)\ge s(\omega)}
\|J\Delta_{\varphi,\psi}^{1/p-1/2}Jh_\omega^{1/2}\|_2\\
&=\inf_{\psi\in \states_*(\Me), s(\psi)\ge s(\omega)}
\|\Delta_{\varphi,\psi}^{1/p-1/2}h_\omega^{1/2}\|_2.
\end{align*}
Assume that $\psi\in \states_*(\Me)$ is such that $h_\omega^{1/2}\in \mathcal D(\Delta_{\varphi,\psi}^{1/p-1/2})$, which
by \cite[Eq. (A.3)]{jencova2018renyi} means that there is some $\eta\in L_2(\Me)s(\psi)$ such that 
$h_\varphi^{1/p-1/2}h_\omega^{1/2}=\eta h_\psi^{1/p-1/2}$ and then $\Delta_{\varphi,\psi}^{1/p-1/2}h_\omega^{1/2}=\eta$. By
Proposition \ref{prop:polar2} (ii) and H\"older inequality, we obtain
\[
\|h_\omega^{1/2}\|_{p,\varphi}^{(2)}=\|h_\varphi^{1/p-1/2}h_\omega^{1/2}\|_p= \|\eta h_\psi^{1/p-1/2}\|_p\le \|\eta\|_2=
\|\Delta_{\varphi,\psi}^{1/p-1/2}h_\omega^{1/2}\|_2.
\]
This shows that $\|h_\omega^{1/2}\|_{p,\varphi}^{(2)}\le \|h_\omega^{1/2}\|_{p,\varphi}^{BST}$. Conversely, let  
$h_\omega^{1/2}=\mu^{1/p}u$ be the $p$-polar decomposition. 
Put $\psi=\mu(1)^{-1}\mu(u\cdot u^*)$, then $\psi\in \states_*(\Me)$,  but note that in general we have  
$s(\psi)=u^*u\le s(\omega)$. Let $\psi_0\in \states_*(\Me)$ 
be any state with $s(\psi_0)= s(\omega)-s(\psi)$ and put
\[
\psi_\epsilon :=\epsilon \psi+(1-\epsilon)\psi_0,\qquad \epsilon\in (0,1).
\]
Then we have $s(\psi_\epsilon)=s(\omega)$. Moreover, 
\[
h_{\psi_\epsilon}^{1/p-1/2}=\epsilon^{1/p-1/2}\mu(1)^{1/2-1/p}u^*h_\mu^{1/p-1/2}u+(1-\epsilon)^{1/p-1/2}h_{\psi_0}^{1/p-1/2}
\]
and
\[
h_\varphi^{1/p-1/2}h_\omega^{1/2}=h_\mu^{1/p}u=k h_{\psi_\epsilon}^{1/p-1/2}
\]
with $k=\epsilon^{1/2-1/p}\mu(1)^{1/p-1/2}h_\mu^{1/2}u$. Hence $h_\omega^{1/2}\in \mathcal
D(\Delta_{\varphi,\psi_\epsilon}^{1/p-1/2})$ and
\[
 \|\xi\|_{p,\varphi}^{BST}\le \|\Delta_{\varphi,\psi_\epsilon}^{1/p-1/2}h_\omega^{1/2}\|_2=\|k\|_2=\epsilon^{1/2-1/p}\mu(1)^{1/p}=\epsilon^{1/2-1/p}\|h_\omega^{1/2}\|_{p,\varphi}^{(2)}.
\]
Letting $\epsilon \to 1$, we obtain the result.

Let $2\le p\le \infty$. Assume that $\|h_\omega^{1/2}\|_{p,\varphi}^{(2)}<\infty$, so that there is some $k=ek\in
L_p(\Me)$ such that $h_\omega^{1/2}=h_\varphi^{1/2-1/p}k$ and $\|h_\omega^{1/2}\|_{p,\varphi}^{(2)}=\|k\|_p$.
Then for any $\psi\in \states_*(\Me)$, we have 
\begin{align*}
\|\Delta_{\psi,\varphi}^{1/2-1/p}h_\omega^{1/2}\|_2&=\|J\Delta_{\psi,\varphi}^{1/2-1/p}Jh_\omega^{1/2}\|_2=
\|\Delta_{\varphi,\psi}^{1/p-1/2}h_\omega^{1/2}\|_2=\|kh_\psi^{1/2-1/p}\|_2\le
\|k\|_p,
\end{align*}
which means that $\|h_\omega^{1/2}\|_{p,\varphi}^{BST}\le \|h_\omega^{1/2}\|_{p,\varphi}^{(2)}$. Next, let 
$\|h_\omega^{1/2}\|_{p,\varphi}^{BST}<\infty$, then we must have $s(\omega)\le s(\varphi)$ and 
$h_\omega^{1/2}=eh_\omega^{1/2}\in \mathcal D(\Delta^{1/2-1/p}_{\psi,\varphi})$ for any $\psi\in \states_*(\Me)$.
For any $\eta\in L_2(\Me)$ with $\|\eta\|_{q,\varphi}^{(2)}=\|\eta\|_{q,\varphi}^{BST}\le 1$, let $\psi\in
\states_*(\Me)$ be such that $s(\psi)\ge s(\omega_\eta)$ and $\eta\in \mathcal D(\Delta^{1/2-1/q}_{\psi,\varphi})$. 
We have
\begin{align*}
|(h_\omega^{1/2},\eta)|&=|(\Delta^{1/2-1/p}_{\psi,\varphi}h_\omega^{1/2},\Delta^{1/2-1/q}_{\psi,\varphi}\eta)|\le
\|\Delta^{1/2-1/p}_{\psi,\varphi}h_\omega^{1/2}\|_2\|\Delta^{1/2-1/q}_{\psi,\varphi}\eta\|_2\\
&\le \|h_\omega^{1/2}\|_{p,\varphi}^{BST}\|\Delta^{1/2-1/q}_{\psi,\varphi}\eta\|_2.
\end{align*}
Since this is true for all such $\psi$, we obtain by Proposition \ref{prop:duality} (ii) 
\[
\|h_\omega^{1/2}\|_{p,\varphi}^{(2)}=\sup_{\eta, \|\eta\|_{q,\varphi}^{(2)}\le 1} |(h_\omega^{1/2},\eta)|
\le \|h_\omega^{1/2}\|_{p,\varphi}^{BST},
\]
this implies the result.

\end{proof}

\section{R\'enyi relative entropies} \label{sec:renyi}

In accordance with \cite{berta2018renyi}, we introduce the following version of the sandwiched R\'enyi relative entropy.

\begin{defi}
For $\psi,\varphi\in  \Me_*^+$ and  $\alpha\in [1/2,1)\cup(1,\infty)$, we define
\begin{equation}\label{eq:BST}
\tilde D_\alpha(\psi\|\varphi):= \frac{1}{\alpha-1}\log \tilde Q_\alpha(\psi\|\varphi)
\end{equation}
where 
\[
\tilde Q_\alpha(\psi\|\varphi):= (\|h^{1/2}_\psi\|^{(2)}_{2\alpha,\varphi})^{2\alpha}.
\]
\end{defi}

By Proposition \ref{prop:bst}, $\tilde D_\alpha$ coincide with the Araki-Masuda divergences defined in \cite{berta2018renyi}.
Moreover, it was proved in \cite[Theorem 3.3]{jencova2018renyi} that for $\alpha>1$, the Araki-Masuda divergences 
 coincide with the
sandwiched R\'enyi relative entropies defined 
in \cite{jencova2018renyi}, so the notation is justified.  The following expression follows easily from Proposition \ref{prop:polar2} (ii).

\begin{thm}\label{thm:sandwiched} Let $\psi\in \Me_*^+$, $\alpha\in [1/2,1)$. Then
\[
\tilde Q_\alpha(\psi\|\varphi)=
\|h_\varphi^{\frac{1-\alpha}{2\alpha}}h_\psi^{1/2}\|^{2\alpha}_{2\alpha}=\Tr(h_\varphi^{\frac{1-\alpha}{2\alpha}}h_\psi
h_\varphi^{\frac{1-\alpha}{2\alpha}})^\alpha .
\]

\end{thm}

It is immediate from the above expression and \cite[Example 2.4]{jencova2018renyi} that $\tilde D_\alpha$ extends the sandwiched R\'enyi relative entropies on density matrices also for
$\alpha\in [1/2,1)$. We also have that for $\lambda,\mu>0$ and all $\alpha\in [1/2,1)\cup (1,\infty]$,
\[
\tilde D_\alpha(\mu\psi\|\lambda\varphi)=\tilde D_\alpha(\psi\|\varphi)+\frac{\alpha}{\alpha-1}\log\mu-\log\lambda.
\]

\subsection{Relation to standard R\'enyi relative entropy}

Recall that the standard R\'enyi relative entropy for $\alpha\in (0,1)$ can be written as
\[
D_\alpha(\psi\|\varphi)=\frac1{\alpha-1}\log (\Tr h_\psi^\alpha h_\varphi^{1-\alpha})=\frac1{\alpha-1}\log \|h_\varphi^{\frac{1-\alpha}2}h_\psi^{\frac{\alpha}2}\|^2_2.
\]
A detailed account on $D_\alpha$ and their properties was  recently given in \cite{hiai2018quantum}.

%

\begin{prop}\label{prop:alt} Let $\varphi,\psi\in \Me_*^+$, $\alpha\in (1/2,1)$.  Then  
\[
\|h_\varphi^{\frac{1-\alpha}2}h_\psi^{\frac{\alpha}2}\|^2_2\le 
\|h_\varphi^{\frac{1-\alpha}{2\alpha}}h_\psi^{1/2}\|_{2\alpha}^{2\alpha}\le 
\psi(1)^{1-\alpha}\|h_\varphi^{\frac{1-\alpha}{2\alpha}}h_\psi^{1-\frac1{2\alpha}}\|_2^{2\alpha}
\]
\end{prop}

\begin{proof}
By H\"older,
\[
\|h_\varphi^{\frac{1-\alpha}{2\alpha}}h_\psi^{1/2}\|_{2\alpha}=\|h_\varphi^{\frac{1-\alpha}{2\alpha}}h_\psi^{1-\frac1{2\alpha}}h_\psi^{\frac{1-\alpha}{2\alpha}}\|_{2\alpha}\le \psi(1)^{\frac{1-\alpha}{2\alpha} }\|h_\varphi^{\frac{1-\alpha}{2\alpha}}h_\psi^{1-\frac1{2\alpha}} \|_2, 
\]
this implies the second inequality. The first inequality was proved in \cite{berta2018renyi}, we add a proof in our
setting. Let us define a function 
\[
f(z)=h_{\varphi}^{1-\alpha z}h_\psi^{\alpha z}=h_{\varphi_0}^{1-\alpha z}eh_\psi^{\alpha z}\in L_1(\Me),\qquad z\in S.
\]
Then $f\in \mathcal F(\Me^R, L_1(\Me))$, so that we can use the properties of the interpolation spaces 
$L_p^R(\Me,\varphi_0)$. Note that  $\|f(1/2)\|_{2,\varphi_0}^R=\|h_\varphi^{\frac{1-\alpha}2}h_\psi^{\frac{\alpha}2}\|_2$. 
Since $1/2=\alpha \frac1{2\alpha}+(1-\alpha)0$, we obtain by Hadamard three lines that
\[
\|f(1/2)\|_{2,\varphi_0}^R\le (\sup_{t\in \mathbb R}\|f(it)\|_{\infty,\varphi_0}^R)^{1-\alpha}
(\sup_{t\in \mathbb R}\|f(\frac 1{2\alpha}+it)\|_{2\alpha,\varphi_0}^R)^\alpha
\]
Let $u_t=h_\varphi^{-i\alpha t}h_\psi^{i\alpha t}$, then $u_t\in \Me$ is a partial isometry, so that 
\[
\|f(it)\|_{\infty,\varphi_0}^R=\|h_\varphi u_t\|_{\infty,\varphi_0}^R=\|u_t\|=1.
\]
Let $\psi_0$ be a faithful state obtained from $\psi$ similarly as $\varphi_0$ from $\varphi$. Then  for $t\in \mathbb R$,
\[
\|f(\frac 1{2\alpha}+it)\|_{2\alpha,\varphi_0}^R=\|h_{\varphi_0}^{-i\alpha t}h_\varphi^{\frac{1-\alpha}{2\alpha}}
h_\psi^{1/2}h_{\psi_0}^{i\alpha t}\|_{2\alpha}=\|h_\varphi^{\frac{1-\alpha}{2\alpha}}
h_\psi^{1/2}\|_{2\alpha},
\]
the last equality holds by \cite[Lemma 10.1]{kosaki1984applications}.

\end{proof}

The next statement is an extension of \cite[Corollary 3.6]{jencova2018renyi} to all values of  $\alpha$. 
Note that the first inequality  for states  of a finite dimensional algebra was proved in \cite[Proposition 11]{wilde2018optimized}. The proof follows easily 
from Proposition \ref{prop:alt} and \cite[Corollary 3.6]{jencova2018renyi}.

\begin{thm}\label{thm:standard} Let $\psi,\varphi\in \states_*(\Me)$ and let $\alpha\in [1/2,1)\cup (1,\infty]$. Then 
\[
D_{2-1/\alpha}(\psi\|\varphi)\le \tilde D_{\alpha}(\psi\|\varphi)\le D_{\alpha}(\psi\|\varphi).
\]

\end{thm}

These inequalities and the limit values for the standard
R\'enyi relative entropies immediately imply that
\[
\lim_{\alpha\uparrow 1} \tilde D_{\alpha}(\psi\|\varphi)=D_1(\psi\|\varphi),
\]
the Araki relative entropy.

\subsection{A variational formula for $\tilde Q_\alpha$}

The next result is an extension of \cite[Lemma 4]{frank2013monotonicity}, obtained in \cite{hiai2020quantum} for $\alpha\in (0,1)$. 
We will use the notation $\Me^{++}$ for the set of 
positive invertible operators in $\Me$.

\begin{prop}[Variational formula] \label{prop:variational} Let $\psi,\varphi\in \Me_*^+$. Then
\begin{enumerate}
\item[(i)] For $\alpha\in (1,\infty)$, we have
\[
\tilde Q_\alpha(\psi\|\varphi)=\sup_{x\in \Me^+} \left( \alpha\Tr h_\psi x -
(\alpha-1)\Tr\left(h_\varphi^{\frac{\alpha-1}{2\alpha}}xh_\varphi^{\frac{\alpha-1}{2\alpha}}\right)^{\frac{\alpha}{\alpha-1}}\right)
\]
\item[(ii)] 
For $\alpha\in [1/2,1)$, we have
\[
\tilde Q_\alpha(\psi\|\varphi)=\inf_{x\in \Me^{++}} \left( \alpha \Tr h_\psi x +
(1-\alpha)\Tr\left(h_\varphi^{\frac{1-\alpha}{2\alpha}}x^{-1}h_\varphi^{\frac{1-\alpha}{2\alpha}}\right)^{\frac{\alpha}{1-\alpha}}\right)
\]
\end{enumerate}

\end{prop}

\begin{proof} Let $\alpha\in (1,\infty)$ and let $\beta=\frac{\alpha}{\alpha-1}$. Assume that $h_\psi\in
L_\alpha(\Me,\varphi)^+$. Recall that  the set $\{h_x=h_\varphi^{1/2}xh_\varphi^{1/2},\ x\in \Me^+\}$ is dense in
$L_p(\Me,\varphi)^+$ for all $p>1$ and we have $\|h_x\|_{p,\varphi}=\|h_\varphi^{1/2p}x
h_\varphi^{1/2p}\|_p$. Recall also the duality pairing $\<\cdot,\cdot\>$ introduced in \cite[Section
2.2]{jencova2018renyi}. We have
\begin{align*}
\sup_{x\in \Me^+} &\left(\alpha\Tr h_\psi x-(\alpha-1)\|h_x\|_{\beta,\varphi}^\beta\right)=\sup_{x\in \Me^+}\left(
\alpha\< h_\psi,
h_x\>-(\alpha-1)\|h_x\|_{\beta,\varphi}^\beta\right)\\
&= \sup_{t\ge 0}\sup_{\substack{x\in \Me^+\\ \|h_x\|_{\beta,\varphi}=t}}
\left(\alpha \<h_x,h_\psi\>-(\alpha-1)t^\beta\right)\\
&=\sup_{t\ge 0} \left(\alpha t\|h_\psi\|_{\alpha,\varphi}-(\alpha-1)t^\beta\right)=\|h_\psi\|_{\alpha,\varphi}^\alpha.
\end{align*}
This proves (i). The statement (ii) was proved in \cite[Lemma 3.18]{hiai2020quantum}.

\end{proof}

It will be useful to introduce the following notations. For $\alpha\in [1/2,1)$ and $\psi,\varphi\in \Me_*^+$, let 
  $\mu_\alpha(\psi\|\varphi)\in \Me_*^+$ be given by 
\begin{equation}\label{eq:mu}
h_{\mu_\alpha(\psi\|\varphi)}:= |h_\varphi^{1/2\alpha-1/2}h_\psi^{1/2}|^{2\alpha}.
\end{equation}
Clearly, if $\mu=\mu_\alpha(\psi\|\varphi)$, then for some partial isometry $u\in \Me$, $h_\psi^{1/2}=\mu^{1/2\alpha}u$  is the $2\alpha$-polar decomposition
with respect to $\varphi$ and by Theorem \ref{thm:sandwiched},
\[
\tilde Q_\alpha(\psi\|\varphi)=\mu_\alpha(\psi\|\varphi)(1).
\]

By \cite[Theorem 4.2]{kosaki1984applicationsuc}, the map $h\mapsto h^{1/p}$ is a homeomorphism of the
positive cones $L_1(\Me)^+\to L_p(\Me)^+$ (we will use this result repeatedly below). By this and
 H\"older inequality, the map $(\psi,\varphi)\mapsto h_\varphi^{1/2\alpha-1/2}h_\psi^{1/2}\in L_{2\alpha}(\Me)$ is jointly
continuous. The continuity of the absolute value $L_p(\Me)\to L_p(\Me)^+$ \cite[Theorem 4.4]{kosaki1984applicationsuc} now implies
 that the map
$\Me_*^+\times \Me_*^+\ni (\psi,\varphi)\mapsto \mu_\alpha(\psi\|\varphi)\in \Me_*^+$
is  jointly (norm) continuous.

 We further put
\begin{equation}\label{eq:xifi}
\xi_{\alpha,\varphi}(x):= h_\varphi^{1/2\gamma} x^{-1} h_\varphi^{1/2\gamma},\qquad x\in \Me^{++},\
\gamma=\frac{\alpha}{1-\alpha}\ge 1
\end{equation}
and
\begin{equation}\label{eq:f}
f_{\alpha,\psi\|\varphi}(x) := \alpha\Tr h_\psi x +(1-\alpha) \|\xi_{\alpha,\varphi}(x)\|_\gamma^\gamma,\qquad x\in
\Me^{++}.
\end{equation}
Then since the function $t\mapsto t^{-1}$ is operator convex, we have
\begin{equation}\label{eq:opconvex}
\xi_{\alpha,\varphi}((1-s)x+sy)\le (1-s)\xi_{\alpha,\varphi}(x)+s\xi_{\alpha,\varphi}(y),\qquad x,y\in \Me^{++},\ s\in
[0,1].
\end{equation}
By the properties of the $L_p$-norms, $x\mapsto f_{\alpha,\psi\|\varphi}$ defines a strictly convex and Fr\'echet
differentiable function $\Me^{++}\to \mathbb R^+$ and by (ii), 
$\tilde Q_\alpha(\psi\|\varphi)=\inf_{x\in \Me^{++}} f_{\alpha,\psi\|\varphi}(x)$.

By the proof of \cite[Lemma 3.18]{hiai2020quantum}, the infimum in (ii) is attained in the special case when there is some 
 $\lambda>0$ such that $\lambda^{-1} \varphi\le \psi\le \lambda\varphi$, this situation will be denoted as $\psi\sim
\varphi$. The next lemma also follows. 

\begin{lemma}\label{lemma:infimum} Let $\psi\sim \varphi$. Then there is some $\bar x\in \Me^{++}$ such that 
\[
\tilde Q_\alpha(\psi\|\varphi)= f_{\alpha,\psi\|\varphi}(\bar x).
\]
Moreover,  we have
\[
\xi_{\alpha,\varphi}(\bar x)=h_{\mu_{\alpha}(\psi\|\varphi)}^{1/\gamma},\qquad  \Tr h_\psi \bar x=
\|h_{\mu_{\alpha}(\psi\|\varphi)}^{1/\gamma}\|_{\gamma}^\gamma=\tilde Q_\alpha(\psi\|\varphi).
\]

\end{lemma}

\section{Data processing inequality}\label{sec:dpi}

The aim of this section  is to prove the following general data processing inequality for $\tilde D_\alpha$ with
$\alpha\in [1/2,1)$. For $\alpha>1$, the DPI was proved in \cite{jencova2018renyi}. 

\begin{thm}[Data processing inequality] \label{thm:DPI_general}
Let $\Phi: L_1(\Me)\to L_1(\Ne)$ be positive and trace preserving. Then for $\alpha\in [1/2,1)$, the DPI holds:
\[
\tilde D_\alpha(\Phi(\psi)\|\Phi(\varphi))\le \tilde D_\alpha(\psi\|\varphi).
\]
\end{thm}

In the case that $\Phi$ is a quantum channel (that is, completely positive and trace preserving), the statement was
proved in \cite{berta2018renyi}. We will give a similar proof here, since it will be used later. The proof for the general
case follows a different strategy, using the variational expression in Proposition \ref{prop:variational}. It will be
presented in Section \ref{sec:dpi_gen}.

\subsection{DPI with respect to quantum channels}

Let $\Phi: L_1(\Me)\to L_1(\mathcal N)$ be a quantum channel. Then the dual map $\Phi^*:\mathcal N\to \Me$ is a completely positive unital normal map. 
Any such map  has a Stinespring representation $(\mathcal K, \pi, T)$, consisting of a Hilbert space $\mathcal K$, a normal *-representation $\pi:\mathcal N\to B(\mathcal K)$ and an isometry $T: L_2(\Me)\to \mathcal K$ such that
\[
\Phi^*(a)=T^*\pi(a)T,\qquad a\in \mathcal N.
\]
Let $\xi\in L_2(\Me)$ be a representing vector for $\psi\in \Me_*^+$, then $T\xi\in \mathcal K$ is a representing vector for $\Phi(\psi)$, hence we have
\[
\tilde D_\alpha(\Phi(\psi)\|\Psi(\varphi))=\frac{2\alpha}{\alpha-1}\log \|T\xi\|^{BST}_{2\alpha,\Phi(\varphi)}.
\]

For $\alpha\in [1/2,1)$, let $\alpha^*>1$ be such that $\frac1{2\alpha}+\frac1{2\alpha^*}=1$. Then the dual parameter
\[
\frac{\alpha^*}{\alpha^*-1}=\frac{\alpha}{1-\alpha}=\gamma.
\]

\begin{thm}\label{thm:DPI_channel} 
Let $\alpha\in [1/2,1)$ and put 
$\mu=\mu_\alpha(\psi\|\varphi)$. Let $\omega\in \Me_*^+$ be such that 
\begin{equation}\label{eq:homega}
h_\omega=h_\varphi^{1/2\gamma}h_\mu^{1/\alpha^*}h_\varphi^{1/2\gamma}.
\end{equation}
Then $\tilde D_{\alpha^*}(\omega\|\varphi)<\infty$ and for any quantum channel $\Phi:L_1(\Me)\to L_1(\Ne)$, we have
\[
\tilde D_\alpha(\psi\|\varphi)\ge \tilde D_\alpha(\Phi(\psi)\|\Phi(\varphi))+ \tilde D_{\alpha^*}(\omega\|\varphi)-
\tilde D_{\alpha^*}(\Phi(\omega)\|\Phi(\varphi))\ge \tilde D_\alpha(\Phi(\psi)\|\Phi(\varphi)).
\]  

\end{thm}

\begin{proof} 
 Put $p=2\alpha$, then $p\in [1,2)$ and the dual parameter
 $q=2\alpha^*\ge 2$. Let $h^{1/2}_\psi=\mu^{1/p}u$ be the $p$-polar decomposition and let 
$\eta= h_\varphi^{1/2-1/q}h_\mu^{1/q}u$, then $\|\eta\|_{q,\varphi}^{(2)}=\mu(1)^{1/q}$ and 
by Proposition \ref{prop:duality} (iii),
\[
\|h_\psi^{1/2}\|_{p,\varphi}^{(2)}= \mu(1)^{-1/q} (\eta,h_\psi^{1/2}).
\]
Let $\Phi: L_1(\Me)\to L_1(\Ne)$ be a quantum channel and let $(\Ka, \pi, T)$ be a Stinespring representation of
$\Phi^*$. Since $T$ is an isometry, we obtain using \eqref{eq:duality_BST}
\[
\|h_\psi^{1/2}\|_{p,\varphi}^{(2)}\|\eta\|_{q,\varphi}^{(2)}=(h_\psi^{1/2},\eta)=\<Th_\psi^{1/2},T\eta\>_{\Ka}\le 
\|Th_\psi^{1/2}\|_{p,\Phi(\varphi)}^{BST}\|T\eta\|_{q,\Phi(\varphi)}^{BST}.
\]
Put $\omega:=\omega_{\eta}\in \Me_*^+$, then $h_\omega =\eta\eta^*$ and \eqref{eq:homega} holds. Moreover, 
$T\eta$ is a vector  representative of $\Phi(\omega)$.
The statement  is now obtained by taking the logarithm of the last inequality, observing that 
$\frac{2\alpha}{\alpha-1}=-\frac{2\alpha^*}{\alpha^*-1}$, and using  DPI for
$\alpha^*>1$.

\end{proof}

\subsection{The proof of general DPI}\label{sec:dpi_gen}

In this paragraph, we will prove Theorem \ref{thm:DPI_general}, using the above variational formula in Proposition 
\ref{prop:variational} (ii). 

Let $\Phi:L_1(\Me)\to L_1(\Ne)$ be a trace preserving positive (not necessarily completely positive) map.
By \cite[Proposition 3.12]{jencova2018renyi}, $\Phi$ restricts to a contraction $L_p(\Me,\varphi)\to
L_p(\Ne,\Phi(\varphi))$ for any $1\le p\le \infty$. Let $i_{p,\varphi}: L_p(\Me)\to L_p(\Me,\varphi)$,
$i_{p,\varphi}(k)=h_\varphi^{1/2q}kh_\varphi^{1/2q}$ be the isometry defined in
\cite[Section 9]{kosaki1984applications}, then we see that 
\[
\Phi_{p,\varphi}= i_{p,\Phi(\varphi)}^{-1}\circ \Phi \circ i_{p,\varphi}
\]
defines a linear contraction $\Phi_{p,\varphi}:L_p(\Me)\to  L_p(\Ne)$.
Note that we have $\Phi_{1,\varphi}=\Phi$ and $\Phi_{\infty,\varphi}=\Phi_\varphi^*$ is the
Petz dual \cite[Section 3.3]{jencova2018renyi}. For any $k\in L_p(\Me)$ and $l\in L_q(\Me,\varphi)$, we have
\[
\<i_{p,\varphi}(k), h_\varphi^{1/2p}lh_\varphi^{1/2p}\>=\Tr kl=\Tr ki_{q,\varphi}^{-1}(h_\varphi^{1/2p}lh_\varphi^{1/2p}),
\]
so that $i_{p,\varphi}^*=i_{q,\varphi}^{-1}$. Since also $\Phi_\varphi$ is the adjoint of $\Phi$ with respect to the duality pairing 
 $\<\cdot,\cdot\>$ between $L_p(\Me,\varphi)$ and $L_q(\Me,\varphi)$, see \cite[Section 2.2]{jencova2018renyi}, 
we see that 
\begin{equation}\label{eq:duality}
\Phi_{p,\varphi}^*=i_{q,\varphi}^{-1}\circ \Phi_\varphi\circ i_{q,\Phi(\varphi)}.
\end{equation}
The following lemma will be useful later.

\begin{lemma}\label{lemma:continuity_petzdual} Let $\Phi:L_1(\Me)\to L_1(\Ne)$ be positive and trace preserving.
Assume that $\varphi_n,\varphi\in \Me_*^+$ are such that $\varphi_n\to \varphi$ (in norm) and  $\Phi(\varphi)$ is
faithful.  Then for any $1\le p\le \infty$ and  $k\in L_q(\Ne)$, we have $ \Phi^*_{p,\varphi_n}(k)\to
\Phi^*_{p,\varphi}(k)$ in $L_q(\Ne)$. 

\end{lemma}

\begin{proof} Let $k\in L_q(\Ne)$. We may assume that 
$k=\Phi(h_\varphi)^{1/2q}x\Phi(h_\varphi)^{1/2q}$ for some $x\in \Ne$, since the set of such elements is dense in
$L_q(\Ne)$ and all the maps are contractions.  In this case, we have
\begin{align}\notag
\Phi_{p,\varphi}^*(k)&= i_{q,\varphi}^{-1}\Phi_\varphi(\Phi(h_\varphi)^{1/2}x\Phi(h_\varphi)^{1/2})=
i_{q,\varphi}^{-1}(h_\varphi^{1/2}\Phi^*(x)h_\varphi^{1/2})\\
&= h_\varphi^{1/2q}\Phi^*(x)h_\varphi^{1/2q}\label{eq:phip}
\end{align}
Let $k_n=\Phi(h_{\varphi_n})^{1/2q}x\Phi(h_{\varphi_n})^{1/2q}$, then $k_n\to k$ and we similarly have
\[
\Phi_{p,\varphi_n}^*(k_n)= h_{\varphi_n}^{1/2q}\Phi^*(x)h_{\varphi_n}^{1/2q}.
\] 
Hence
\begin{align*}
\|\Phi_{p,\varphi_n}^*(k)-\Phi_{p,\varphi}^*(k)\|_q&\le \|k-k_n\|_q
+\|\Phi_{p,\varphi_n}^*(k_n)-\Phi_{p,\varphi}^*(k)\|_q\\
&\le \|k-k_n\|_q+ \|
h_{\varphi_n}^{1/2q}\Phi^*(x)h_{\varphi_n}^{1/2q}-h_{\varphi}^{1/2q}\Phi^*(x)h_{\varphi}^{1/2q}\|_q\\
&\to 0
\end{align*}

\end{proof}

We will also need the following inequality due to Kosaki, see \cite[Lemma 3.3]{kosaki1984applicationsuc}:
\begin{equation}\label{eq:kosaki_order}
\|k-l\|_p^p\le \|k\|_p^p-\|l\|_p^p,\qquad 0\le l\le k\in L_p(\Me).
\end{equation}

\begin{proof}[Proof of Theorem \ref{thm:DPI_general}]
Clearly, the DPI is equivalent to $ \tilde Q_\alpha(\Phi(\psi)\|\Phi(\varphi))\ge \tilde Q_\alpha(\psi\|\varphi)$.
Let $y\in \Ne^{++}$. Since $\Phi^*$ is positive and unital, we have  by the Choi inequality (see \cite[Corollary
2.3]{choi1974aschwarz}) that
$\Phi^*(y)^{-1}\le \Phi^*(y^{-1})$, 
so that 
\begin{align}
\xi_{\alpha,\varphi}(\Phi^*(y))&=h_\varphi^{1/2\gamma}\Phi^*(y)^{-1}h_\varphi^{1/2\gamma}\le 
h_\varphi^{1/2\gamma}\Phi^*(y^{-1})h_\varphi^{1/2\gamma}\notag\\
&=\Phi_{\alpha^*,\varphi}^*(\Phi(h_\varphi)^{1/2\gamma}y^{-1}\Phi(h_\varphi)^{1/2\gamma})=
\Phi_{\alpha^*,\varphi}^*(\xi_{\alpha,\Phi(\varphi)}(y))\label{eq:xineq_Phi},
\end{align}
here the second equality is obtained as in  \eqref{eq:phip}.
By \eqref{eq:kosaki_order}, this implies 
\begin{equation}\label{eq:xineq}
\|\xi_{\alpha,\varphi}(\Phi^*(y))\|_{\gamma}^\gamma\le
\|\Phi_{\alpha^*,\varphi}^*(\xi_{\alpha,\Phi(\varphi)}(y))\|_{\gamma}^\gamma\le
\|\xi_{\alpha,\Phi(\varphi)}(y)\|_\gamma^\gamma,
\end{equation}
the last inequality follows since $\Phi^*_{\alpha^*,\varphi}$ is a contraction.  Putting all together, we obtain
 for any $y\in \Ne^{++}$ that
\begin{align}
\tilde Q_\alpha(\psi\|\varphi)&=\inf_{x\in \Me^{++}} \left( \alpha \Tr h_\psi x +
(1-\alpha)\|\xi_{\alpha,\varphi}(x)\|_{\gamma}^\gamma\right)\label{eq:DPI_1}\\
&\le \alpha \Tr h_\psi \Phi^*(y)+(1-\alpha)\|\xi_{\alpha,\varphi}(\Phi^*(y))\|_{\gamma}^\gamma\\
&\le \alpha \Tr \Phi(h_\psi)y +(1-\alpha)\|\xi_{\alpha,\Phi(\varphi)}(y)\|_{\gamma}^\gamma\\
&= f_{\alpha,\Phi(\psi)\|\Phi(\varphi)}(y)\label{eq:DPI_last}
\end{align}
which implies that 
\[
\tilde Q_\alpha(\psi\|\varphi)\le \tilde Q_\alpha(\Phi(\psi)\|\Phi(\varphi)).
\]
 
\end{proof}

\section{Equality conditions and sufficiency}\label{sec:suffic}

In this section we prove the following extension of  \cite[Theorem 4.6]{jencova2018renyi}.

\begin{thm}\label{thm:suffic}
Let $\alpha\in (1/2,1)$ and assume that the linear map $\Phi: L_1(\Me)\to L_1(\Ne)$ is 2-positive and trace preserving.
Let $\psi,\varphi\in \Me_*^+$ be such that  $s(\psi)\le s(\varphi)$. Then the equality
\[
\tilde D_\alpha(\psi\|\varphi)=\tilde D_\alpha(\Phi(\psi)\|\Phi(\varphi))
\]
holds if and only if  $\Phi$ is sufficient with respect to $\{\psi,\varphi\}$. 

\end{thm}

Because of the assumption on the supports, we may and will suppose as in the proof of \cite[Theorem
4.6]{jencova2018renyi} that both $\varphi$ and $\Phi(\varphi)$ are
faithful. 

As in Section \ref{sec:dpi}, we first provide a proof in the case that $\Phi$ is a quantum channel.

\begin{proof}[Proof for quantum channels]  Let $\alpha^*$, $\mu$ and $\omega$ be as in Theorem \ref{thm:DPI_channel}. 
Assume that the equality holds,  then we must have 
\[
\tilde D_{\alpha^*}(\omega\|\varphi)=\tilde D_{\alpha^*}(\Phi(\omega)\|\Phi(\varphi)).
\]
Since $1<\alpha^*<\infty$, this equality implies that $\Phi$ is sufficient with respect to $\{\omega,\varphi\}$
 (\cite[Theorem 4.6]{jencova2018renyi}). By \eqref{eq:homega} and \cite[Lemma 4.4]{jencova2018renyi}, $\Phi$ is sufficient with respect to $\{\mu(1)^{-1}\mu,\varphi\}$.

Let $E:\Me\to\Me$ be a faithful normal conditional expectation as in \cite[Lemma 4.3]{jencova2018renyi}, so that
$\varphi\circ E=\varphi$ and for any $\rho\in \states_*(\Me)$,  $\Phi$ is sufficient with respect to $\{\rho,\varphi\}$
if and only if $\rho\circ E=\rho$. Let $p=2\alpha$ and let $E_{p}$ be the extension of $E$ to $L_{p}(\Me)$ 
(\cite{junge2003noncommutative}, \cite[Appendix A.3]{jencova2018renyi}).  Similarly as in the proof of \cite[Lemma
4.4]{jencova2018renyi}, we have $E_p(h_\mu^{1/p})=h_\mu^{1/p}$ and by \cite[Eq. (A.7)]{jencova2018renyi},
\[
h_\varphi^{1/p-1/2}h_\psi^{1/2}u^*=h_\mu^{1/p}=E_p(h_\mu^{1/p})=h_\varphi^{1/p-1/2}E_2(h_\psi^{1/2}u^*).
\]
Since $\varphi$ is faithful, we have $u^*u\ge s(\psi)$ and the above equalities imply that $h_\psi^{1/2}u^*=E_2(h_\psi^{1/2}u^*)$. 
Hence
\[
h_{\psi\circ E}=E_1(h_\psi)=h_\psi^{1/2}u^*uh_\psi^{1/2}=h_\psi
\]
so that $\Phi$ is sufficient for $\{\psi,\varphi\}$. The converse is obvious from DPI.

\end{proof}

We next prove Theorem \ref{thm:suffic} in the case $\psi\sim\varphi$. Note that this holds e.g. 
if $\Me$ is finite dimensional and
$s(\psi)=s(\varphi)$. 

\begin{proof}[Proof for $\psi\sim\varphi$] By Lemma \ref{lemma:infimum}, there are some $\bar x\in \Me^{++}$ and $\bar
y\in \Ne^{++}$ such that $\tilde Q_\alpha(\psi\|\varphi)=f_{\alpha,\psi\|\varphi}(\bar x)$ and 
$\tilde Q_\alpha(\Phi(\psi)\|\Phi(\varphi))=f_{\alpha,\Phi(\psi)\|\Phi(\varphi)}(\bar y)$. The equality in DPI implies
that with $y=\bar y$, all the inequalities between \eqref{eq:DPI_1} and \eqref{eq:DPI_last}, and hence also in
\eqref{eq:xineq}, must be equalities. Since $f_{\alpha,\psi\|\varphi}$ is strictly convex, the infimum is attained an a
unique point and hence $\bar x=\Phi^*(\bar y)$. Using \eqref{eq:xineq_Phi}, equality in the first inequality of \eqref{eq:xineq} and \eqref{eq:kosaki_order}, we obtain
\[
\xi_{\alpha,\varphi}(\bar x)=\xi_{\alpha,\varphi}(\Phi^*(\bar
y))=\Phi^*_{\alpha^*,\varphi}(\xi_{\alpha,\Phi(\varphi)}(\bar y)).
\]
By Lemma \ref{lemma:infimum}, this means that $h_\mu^{1/\gamma}=\Phi^*_{\alpha^*,\varphi}(h_\nu^{1/\gamma})$, where 
$\mu=\mu_\alpha(\psi\|\varphi)$ and $\nu=\mu_\alpha(\Phi(\psi)\|\Phi(\varphi))$. By the equality in
DPI, we obtain
\[
\|h_\nu^{1/\gamma}\|_{\gamma}^\gamma=\nu(1)=\tilde Q_\alpha(\Phi(\psi)\|\Phi(\varphi))=\tilde Q_\alpha(\psi\|\varphi)=
\|h_\mu^{1/\gamma}\|_\gamma^\gamma=\|\Phi^*_{\alpha^*,\varphi}(h_\nu^{1/\gamma})\|_\gamma^\gamma.
\]

Let now $\sigma\in \Me_*^+$ be such that $h_\sigma= h_\varphi^{1/2\alpha^*}h_\mu^{1/\gamma}h_\varphi^{1/2\alpha^*}$ 
and $\rho\in \Ne_*^+$ be such that $h_\rho=\Phi(h_\varphi)^{1/2\alpha^*}h_\nu^{1/\gamma}\Phi(h_\varphi)^{1/2\alpha^*}$.
By \eqref{eq:duality} and the equalities above, we obtain 
\[
\Phi_\varphi(h_\rho)=
h_\varphi^{1/2\alpha^*}\Phi_{\alpha^*,\varphi}^*(h_{\nu}^{1/\gamma})h_\varphi^{1/2\alpha^*}=h_\varphi^{1/2\alpha^*}h_\mu^{1/\gamma}h_\varphi^{1/2\alpha^*}=
h_\sigma
\]
Further,
\[
\|h_\rho\|_{\gamma,\Phi(\varphi)}=\|h_\nu^{1/\gamma}\|_\gamma=\|\Phi_{\alpha^*,\varphi}^*(h_{\nu}^{1/\gamma})\|_{\gamma}=\|\Phi_\varphi(h_\rho)\|_{\gamma,\varphi}.
\]
By \cite[Theorem 4.6]{jencova2018renyi}, this implies  that the 2-positive trace preserving map 
$\Phi_\varphi: L_1(\Ne)\to L_1(\Me)$ is sufficient with respect to $\{\rho,\Phi(\varphi)\}$. Since the Petz dual of
$\Phi_\varphi$ with respect to $\Phi(\varphi)$ is again $\Phi$, we obtain that $\Phi\circ \Phi_\varphi(\rho)=\rho$.
 Putting all together, we get
\[
\Phi_\varphi\circ \Phi(\sigma)=\Phi_\varphi\circ \Phi\circ \Phi_\varphi(\rho)=\Phi_\varphi(\rho)=\sigma.
\]
Hence $\Phi$ is sufficient with respect to $\{\sigma,\varphi\}$. The proof is finished similarly as the proof for
quantum channels above.

\end{proof}

\subsection{The general case}

In the general case, we need some refinement of the inequalities \eqref{eq:DPI_1} - \eqref{eq:DPI_last}. For this, we
will use the fact that the spaces $L_p(\Me)$, $1<p<\infty$ are uniformly convex.
Throughout this section, we fix the notations 
\[
\mu=\mu_{\alpha}(\psi\|\varphi),\quad \nu=\mu_{\alpha}(\Phi(\psi)\|\Phi(\varphi)),\quad f=f_{\alpha,\psi\|\varphi},\quad 
\xi=\xi_{\alpha,\varphi}.
\]

\begin{lemma}\label{lemma:f_ineq} Let $\psi\sim \varphi$. Then for any $y\in \Me^{++}$, we have
\[
f(y)-\tilde Q_\alpha(\psi\|\varphi)\ge
2(1-\alpha)\left[\frac12\|h^{1/\gamma}_\mu\|_\gamma^\gamma+\frac12\|\xi(y)\|_\gamma^\gamma-\|\frac12(h_\mu^{1/\gamma}+\xi(y))\|_\gamma^\gamma\right]>0.
\]
In particular, for any $\epsilon >0$, there is some $\delta>0$ (independent from $\varphi$ and $\psi$) such that 
 if $\|\xi(y)-h_\mu^{1/\gamma}\|_\gamma\ge \epsilon$, then 
\[
f(y)-\tilde Q_\alpha(\psi\|\varphi)\ge
\delta(\|h^{1/\gamma}_\mu\|_\gamma^\gamma+\|\xi(y)\|_\gamma^\gamma).
\]

\end{lemma}

\begin{proof} 
For $x,y\in \Me^{++}$, we have
\begin{align*}
\<\nabla f(x),y-x\>=&\lim_{s\to 0^+} s^{-1}[ f((1-s)x+sy)-f(x)]\\
=&\alpha \Tr(h_\psi(y-x))\\&+(1-\alpha)\lim_{s\to 0^+}s^{-1}
\left[\|\xi((1-s)x+sy)\|_{\gamma}^\gamma-\|\xi(x)\|_{\gamma}^\gamma\right].
\end{align*}
For $s\in (0,1/2)$, we have by \eqref{eq:opconvex} and \eqref{eq:kosaki_order}
\begin{align*}
\|\xi((1-s)x+sy)\|_{\gamma}^\gamma&\le
\|(1-s)\xi(x)+s\xi(y)\|_{\gamma}^\gamma\\
&=\|(1-2s)\xi(x)+2s
\frac12(\xi(x)+\xi(y))\|_{\gamma}^\gamma\\
&\le (1-2s)\|\xi(x)\|_{\gamma}^\gamma+2s\|\frac12(\xi(x)+\xi(y))\|_{\gamma}^\gamma
\end{align*}
 Using this in the above computation, we get
\begin{align*}
\<\nabla f(x),y-x\>\le
f(y)&-f(x)\\-&2(1-\alpha)\left[\frac12\|\xi(x)\|_\gamma^\gamma+\frac12\|\xi(y)\|_\gamma^\gamma-\|\frac12
(\xi(x)+\xi(y))\|_\gamma^\gamma\right].
\end{align*}
By Lemma \ref{lemma:infimum}, there is some $\bar x\in \Me^{++}$ such that 
$ \tilde Q_\alpha(\psi\|\varphi)=f(\bar x)$, in this case we have $\nabla f(\bar x)=0$ and $\xi(\bar
x)=h_{\mu}^{1/\gamma}$, this yields the first part of the lemma. The second part follows by uniform
convexity of $L_\gamma(\Me)$ (see e.g. \cite{diestel1975geometry}).

\end{proof}

\begin{lemma}\label{lemma:QPhi_lower} Let $\psi\sim \varphi$ and let $\Phi: L_1(\Me)\to L_1(\Ne)$ be positive and trace preserving. 
Let $\bar y\in \Ne^{++}$ be such that $\tilde Q_\alpha(\Phi(\psi)\|\Phi(\varphi))= f_{\alpha,\Phi(\psi)\|\Phi(\varphi)}(\bar
y)$. 
Then we have
\[
\tilde Q_\alpha(\Phi(\psi)\|\Phi(\varphi))\ge f(\Phi^*(\bar
y))+(1-\alpha)\|\Phi_{\alpha^*,\varphi}^*(h_\nu^{1/\gamma})-\xi(\Phi^*(\bar y))\|_\gamma^\gamma
\]

\end{lemma}

\begin{proof} Since $\xi_{\alpha,\Phi(\varphi)}(\bar y)=h_\nu^{1/\gamma}$, 
we have 
\begin{align*}
\tilde Q_\alpha(\Phi(\psi)\|\Phi(\varphi))&=f(\Phi^*(\bar y))
+ (1-\alpha)\left[\|h_\nu^{1/\gamma}\|_\gamma^\gamma-\|\xi(\Phi^*(\bar y))\|_\gamma^\gamma\right]\\
&\ge f(\Phi^*(\bar y))
+ (1-\alpha)\left[\|\Phi_{\alpha^*,\varphi}^*(h_\nu^{1/\gamma})\|_\gamma^\gamma-\|\xi(\Phi^*(\bar
y))\|_\gamma^\gamma\right].
\end{align*}
We obtain the result by \eqref{eq:xineq_Phi} and  \eqref{eq:kosaki_order}. 
\end{proof}

Now we can prove Theorem \ref{thm:suffic} in the general case.

\begin{proof}[Proof of Theorem \ref{thm:suffic}] Fix some  sequences $\varphi_n\to \varphi$ and $\psi_n\to \psi$ in
$\Me_*^+$ such that $\psi_n\sim \varphi_n$. By joint continuity of $\tilde Q_\alpha$ \cite[Theorem 3.15 (3)]{hiai2020quantum} and the assumption, we have 
\[
\lim_n\tilde Q_\alpha(\psi_n\|\varphi_n)=\lim_n Q_\alpha(\Phi(\psi_n)\|\Phi(\varphi_n))= \tilde Q_\alpha(\psi\|\varphi)
\]

Put
\[
\mu_n=\mu_\alpha(\psi_n\|\varphi_n),\quad \nu_n=\mu_\alpha(\Phi(\psi_n)\|\Phi(\varphi_n)), \quad f_n=f_{\alpha,\psi_n\|\varphi_n}.
\]
We have $\mu_n\to \mu$ and $\nu_n\to \nu$, so that  $h_{\mu_n}^{1/\gamma}\to
h_\mu^{1/\gamma}$ and $h_{\nu_n}^{1/\gamma}\to h_\nu^{1/\gamma}$. By Lemma \ref{lemma:continuity_petzdual}, we obtain 
$\Phi_{\alpha^*,\varphi_n}^*(h_{\nu_n}^{1/\gamma})\to \Phi_{\alpha^*,\varphi}^*(h_{\nu}^{1/\gamma})$.   

Let  $y_n\in \Ne^{++}$ be such that 
$\tilde Q_\alpha(\Phi(\psi_n)\|\Phi(\varphi_n))=f_{\alpha,\Phi(\psi_n)\|\Phi(\varphi_n)}(y_n)$ and
let us denote
$\xi_n=\xi_{\alpha,\varphi_n}(\Phi^*(y_n))$. Using Lemma \ref{lemma:QPhi_lower}, we get
\begin{align}
\tilde Q_\alpha(\Phi(\psi_n)\|\Phi(\varphi_n))-\tilde Q_\alpha(\psi_n\|\varphi_n)\ge &f_n(\Phi^*(y_n))
-\tilde Q_\alpha(\psi_n\|\varphi_n)\label{eq:proof1} \\
&+ (1-\alpha)\|\Phi_{\alpha^*,\varphi_n}^*(h_{\nu_n}^{1/\beta})-\xi_n\|_\beta^\beta\label{eq:proof2}
\end{align}
 Since both
quantities on the right hand side of \eqref{eq:proof1} and \eqref{eq:proof2}  are nonnegative, we immediately obtain that 
$\xi_n\to \Phi_{\alpha^*,\varphi}^*(h_{\nu}^{1/\gamma})$ and by Lemma \ref{lemma:f_ineq}, we also get $\xi_n\to
h_\mu^{1/\gamma}$. This proves  $\Phi_{\alpha^*,\varphi}^*(h_{\nu}^{1/\gamma})=h_\mu^{1/\gamma}$.

 The proof now can be finished exactly as in the case $\psi\sim \varphi$.

\end{proof}

\begin{rem} In the case $\gamma\ge 2$ (that is, $\alpha\ge 2/3$),  we may use the Clarkson inequality
\cite[Theorem 6.6]{kosaki1984applicationsuc} in Lemma \ref{lemma:f_ineq} and obtain
\[
f(y)-\tilde Q_\alpha(\psi\|\varphi)\ge (1-\alpha)2^{1-\gamma}\|\xi(y)-h_\mu^{1/\gamma}\|_\gamma^\gamma.
\]
Using this with $y=\Phi^*(\bar y)$, we get from Lemma \ref{lemma:QPhi_lower} that
\begin{align*}
\tilde Q_\alpha&(\Phi(\psi)\|\Phi(\varphi))-\tilde Q_\alpha(\psi\|\varphi)\\&\ge (1-\alpha)2^{1-\gamma}\left[\|\xi(\Phi^*(\bar
y))-h_\mu^{1/\gamma}\|_\gamma^\gamma+\|\Phi^*_{\alpha^*,\varphi}(h_\nu^{1/\gamma})-\xi(\Phi^*(\bar
y))\|_\gamma^\gamma\right]\\
&\ge (1-\alpha)4^{1-\gamma}\|\Phi^*_{\alpha^*,\varphi}(h_\nu^{1/\gamma})-h_\mu^{1/\gamma}\|_\gamma^\gamma.
\end{align*}
By a limit argument, this inequality holds for all $\psi,\varphi$ with  $s(\psi)\le s(\varphi)$ and positive trace
preserving maps $\Phi$. Another DPI lower bound (for all values of $\alpha\in [1/2,1)$) is in Lemma \ref{lemma:dpi_lower}
 below.

\end{rem}

\begin{lemma}\label{lemma:dpi_lower} Let $\psi,\varphi\in \Me_*^+$ with  $s(\psi)\le s(\varphi)$ and let 
$\Phi: L_1(\Me)\to L_1(\Ne)$ be positive and trace
preserving. Then we have
\[
\tilde Q_\alpha(\Phi(\psi)\|\Phi(\varphi))-\tilde Q_\alpha(\psi\|\varphi)\ge (1-\alpha)\left[
\|h_\nu^{1/\gamma}\|_{\gamma}^\gamma-\|\Phi^*_{\alpha^*,\varphi}(h_\nu^{1/\gamma})\|_{\gamma}^\gamma\right].
\]

\end{lemma}

\begin{proof} Assume first that $\psi\sim \varphi$ and let $\bar y\in \Ne^{++}$ be chosen as before.
Then by Lemma \ref{lemma:infimum}
\begin{align*}
\tilde Q_\alpha(\psi\|\varphi)&\le \alpha\Tr \Phi(\psi)\bar y+
(1-\alpha)\|\Phi^*_{\alpha^*,\varphi}(h_\nu^{1/\gamma})\|_{\gamma}^\gamma\\
&= \alpha \tilde Q_\alpha(\Phi(\psi)\|\Phi(\varphi)) +(1-\alpha)\|\Phi^*_{\alpha^*,\varphi}(h_\nu^{1/\gamma})\|_{\gamma}^\gamma.
\end{align*}

The general case is proved by a limit argument.

\end{proof}

\section*{Acknowledgements}  I am indebted to Fumio Hiai for sharing the manuscript of his monograph
\cite{hiai2020quantum} and useful discussions and comments. His variational formula and its proof inspired a large part
of this paper. The research was supported by the grants APVV-16-0073 and VEGA 2/0142/20.


\end{document}